\documentstyle[multicol,aps,prb,psfig,amstex,amssymb]{revtex}

\begin{document}


\title{Current facilitation by plasmon resonances between parallel wires of finite length}

\author{A. L. Chudnovskiy \\
I Institut f\"ur Theoretische Physik,  Universit\"at Hamburg,  \\
Jungiusstr. 9, D-20355 Hamburg, Germany}

\date{\today}


\maketitle

\begin{abstract}
The current voltage (IV) characteristics for perpendicular transport through two sequentially
coupled wires of finite length is calculated analytically.
The transport within a Coulomb blockade step is assisted by plasmon resonances that appear as
steps in the IV characteristics with positions and heights
depending on inter- and intrawire interactions. In particular, due to the interwire interactions,
the peak positions shift to lower voltages in comparison to the noninteracting wires
which reflects the facilitation  of current by interactions. The interwire interactions are also
found to enhance the thermally activated current.
\end{abstract}

Recent technological progress in creation of clean one-dimensional wires, either as
carbon nanotubes \cite{nanotubes} or as a part of semiconducting structures
\cite{Auslaender}
has attracted much attention to the study of transport through
quasi-one-dimensional systems, both experimentally and theoretically.
\cite{nanotubes,Auslaender,tubes2,Halperin,Boese,Governale,Sassetti} Transport through
clean quasi-one dimensional systems probes the main consequences of the
Luttinger liquid  (LL) theory \cite{Haldane}, the spin-charge separation and
the charge fractionalization. Experimentally, quasi-1d systems have a finite length, or
they are effectively divided into finite size fragments by impurities \cite{Sassetti}.
If the length is not very large, the quasi-1d systems actually represent one-dimensional
quantum dots, with finite charging energy and discrete spectrum. \cite{nanotubes,nanofin}
Therefore, the theoretical study of finite-size effects on transport
through wires of finite length $L$ that form the LL in the limit $L\rightarrow\infty$ is of
much importance for the interpretation of experiments \cite{Sassetti}.

\begin{figure}[p]
\psfig{file=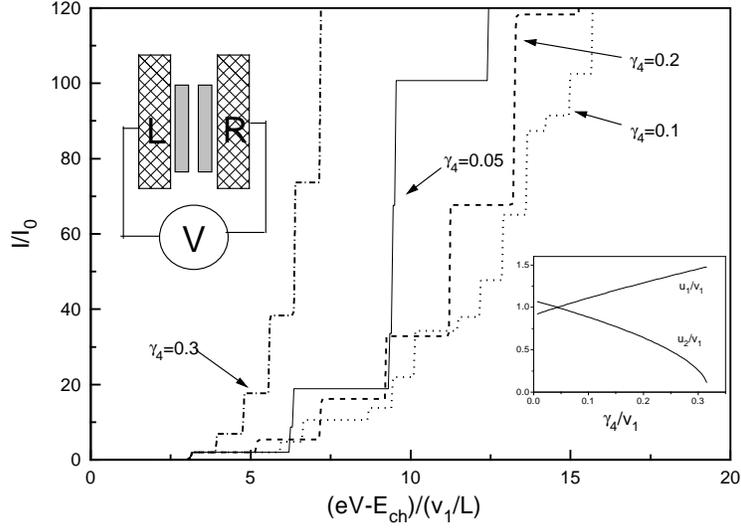,width=17cm,height=12cm,angle=-90}
\vskip -2cm
\caption{IV characteristics for different interwire interaction constants $\gamma_4$.
The position of the second step reflecting the current through the first excited plasmon
mode shifts to lower voltages with growing $\gamma_4$, which indicates the facilitation of
current by interactions. In the shown voltage range, the curves with $\gamma_4=0.2;0.3$
show equidistant steps, reflecting the contribution of the mode $u_2$ only,
whereas both modes $u_1$ and $u_2$ contribute to the curve with $\gamma_4=0.1$.
For $\gamma_4=0.5$, $u_1\approx u_2$, the positions of the
steps from both modes coincide, the height of each step is the sum of the two contributions.
Normalization of the current $I_0=16\pi e|T_{i,r}|^2\rho_r(E_F)/a$.
Inset: the velocities of plasmon modes $u_1$ and $u_2$ vs $\gamma_4$.
Other parameters: $g_1=0.28$, $g_2=0.26$, $\gamma_2=0.05$.}
\label{IV-d}
\end{figure}

In this Letter, we address those features of a LL that can be seen in experiments on
perpendicular transport through  two coupled
quantum wires of finite length $L$, analyzing the current--voltage (IV)  characteristics
of the circuit consisting of the left reservoir, two quantum wires, and the right reservoir,
coupled in sequence (see the inset in Fig. \ref{IV-d}). The reservoirs are assumed to be noninteracting.
There is a magnetic field  applied perpendicular to the plane of the wires.
Such system can be realized in form of two carbon nanotubes. A similar geometry has been
employed in experiments on semiconductor structures  \cite{Auslaender}
with the difference that only one of the wires is of finite length there.
Since the discreteness of the spectrum is explicitely taken into account, the results
presented here may find analogies in experiments on double quantum
dots \cite{Holleitner}.
We do not consider spin effects, the Luttinger liquids are assumed to be spinless.
We derive an analytical expression for the current-voltage characteristics in lowest order
in tunneling to the reservoirs, concentrating on the region within a single Coulomb blockade
step.

We found that  the interwire interactions facilitate
transport at finite bias voltages, as well as thermally activated transport.
Our results on IV characteristics  are summarized in  Fig. \ref{IV-d}.
The discreteness of the spectrum of a finite-length wire transforms the power-law singularities,
typical for the LL, into a series of discrete steps with envelope described by the power-law
function of the infinite system \cite{Zuelicke}. Each step in the IV characteristics corresponds
to a plasmon resonance between the LL wires.
The {\it positions} of the steps of the IV curve reflect the
quantization of the {\it charge-mode} of the LL, in particular,
the energies of two independent bosonic modes, propagating on the length $L$ with velocities
$u_1$ and $u_2$.\cite{Governale} One of the velocities ($u_2$ in our case) always renormalizes
down by interwire interactions, which results in the lowering of plasmon excitation energies,
and hence the facilitation of transport through excited states as compared to the noninteracting
wires.
The height of each step reflects the strength of the corresponding plasmon resonance, and
is also controlled by LL  interaction constants.
Therefore, IV characteristics  should clearly demonstrate the
effect of LL interactions. The outlined features are rather insensitive to the geometry of the
reservoirs. They can be seen in the case of  two- and three-dimensional reservoirs contacted
along a whole wire, as well as for a circuit with  point-like coupling between
the reservoirs and the wires, such as that realized in STM experiments \cite{nanotubes,tubes2}.
In contrast, due to strong restrictions imposed by momentum conservation by tunneling,
the interaction effects can hardly be seen in the case of one-dimensional reservoirs parallel to
the wires.

Our starting point is the Hamiltonian of two interacting
Luttinger liquids of finite length, with different Fermi-velocities $v_1$ and $v_2$ and  with a
single-particle (1P) tunneling of amplitude $t$ between the LL's. The length of the wires $L$ is
assumed to be large,
so that the wave number $k$ is conserved in the tunneling process.   In the presence of a magnetic field $B$ perpendicular to the
plane, the states of the wire 1 and the left reservoir acquire an additional phase factor
$e^{ibx}$  ($b=eB/\hbar$, $x$ denotes a coordinate along the wire) with respect to the states of
the wire 2 and the right reservoir \cite{Governale}. The application of the LL approach to wires
of finite length is
limited by the condition that the screening length for the Coulomb interaction in the wire is much
less than the length of the wire $L$. This condition is fulfilled, for example, in recent
experiments on semiconductors \cite{Auslaender,Halperin}.

We take into account the electrostatic charging energy phenomenologically,
shifting the groundstate energy by the capacitive energy
$E(N_1,N_2)$, where $N_1,N_2$ denote the total number of electrons in the first and the second
wire respectively. The calculation of $E(N_1,N_2)$ has been performed
previously. \cite{Boese,Pfannkuche} Furthermore, it is assumed that the application of
voltage between the wires shifts the
energy of the one-particle states of the two wires by the amount $eV$ relative to each other
(band-shifting case \cite{Boese}).

Since transport between the LL's is determined by plasmon resonances, and in resonance the
interwire tunneling has to be treated exactly,  we first diagonalize the 1P
spectrum.
The spectrum of the diagonalized 1P Hamiltonian consists of two branches with dispersions
$\epsilon_{\pm}(k)=\left[eV+(v_1+v_2)k\pm\sqrt{(eV+(v_1-v_2)k)^2+4t^2}
\right]/2$.
In general, the resulting spectrum is not linear any more for $v_1\neq v_2$. However,
as long as
$|eV(v_1-v_2)k|\ll eV^2+4t^2$
one can expand the square root, and retaining only the
first term of expansion consider the spectrum as approximately linear.
Further we work in that approximation.
Upon inclusion of two-particle interactions, the two branches $\epsilon_{\pm}(k)$
form two new interacting Luttinger liquids.

After the diagonalization of the 1P part of the two-wire Hamiltonian, the physical picture
changes from the sequential circuit, the left reservoir -- two coupled wires -- the right
reservoir, to the circuit, where there are two LL's coupled in parallel to the left and
right reservoirs with no direct tunneling between the Luttinger liquids.

Note that {\it two particle} hopping terms between the new LL's in the
particle-particle ($pp$)  and particle-hole ($ph$) channels \cite{Finkelstein} are generated
by the diagonalization of 1P Hamiltonian.
These terms will be omitted in further consideration. Their amplitudes are small at small
tunneling $t$, though they play a crucial role close to the instability in the
$ph$ or $pp$ channel (see below).

After the 1P diagonalization, the Hamiltonian of
interacting LL's {\it without} the one-particle tunneling is of the form
$
H=\sum_{\nu=1,2}H_{\nu}+H_{12},
$
where $H_{\nu}$ is a standard LL Hamiltonian without backward scattering \cite{HS-JvD},
and the interwire interaction is of the form
$
H_{12}=v_1\int_{-L/2}^{L/2}\frac{dx}{2\pi}:\left[\gamma_2\hat{\rho}_{1L}
\hat{\rho}_{2R}+
\gamma_4\hat{\rho}_{1L}\hat{\rho}_{2L}+ L\leftrightarrow R
\right]:.
$
$\gamma_2$ and $\gamma_4$ are the interwire interaction constants,
$\hat{\rho}^\nu_{R,L}(x)$ denote the chiral electron density operators of the two wires.

In the bosonized form, the Hamiltonian looks like
(here $\vec\Phi\equiv\left(\Phi_1,\Phi_2\right)^T$,
$\vec\Theta\equiv\left(\Theta_1,\Theta_2\right)^T$)
\begin{equation}
H=\frac{1}{4}\int_{-L/2}^{L/2}\frac{dx}{2\pi}\left\{:\left(\partial_x\vec\Phi^T\right)
\hat{M}_{\phi}\left(\partial_x\vec\Phi\right):+
:\left(\partial_x\vec\Theta^T\right)
\hat{M}_{\theta}\left(\partial_x\vec\Theta\right):\right\},
\label{Hphth}
\end{equation}
where
$
\hat{M}_{\phi}=\left(
\begin{array}{cc}
v_1g_1  & v_1(\gamma_4 -\gamma_2) \\
v_1(\gamma_4-\gamma_2) & v_2g_2
\end{array}
\right)
$,
$
\hat{M}_{\theta}=\left(
\begin{array}{cc}
v_1/g_1  & v_1(\gamma_4+\gamma_2) \\
v_1(\gamma_4+\gamma_2) & v_2/g_2
\end{array}
\right)
$, $v_\nu$ is the charge mode velocity in  wire $\nu$ and $g_\nu$ is the
LL interaction constant for wire $\nu$.
Diagonalization of the Hamiltonian (\ref{Hphth}) is achieved by applying a composition of
unitary rotations and subsequent rescalings, whereby the rescalings preserve the duality of
the fields $\Phi_i$ and $\Theta_i$: $\Phi_i\rightarrow\Phi_i\sqrt{K_i}$,
$\Theta_i\rightarrow\Theta_i/\sqrt{K_i}$. The same has been achieved by other means in
\cite{Governale,Finkelstein,Shahbazyan}.
The diagonalization renders (\ref{Hphth}) to the
Hamiltonian of two uncoupled LL's with {\it different} charge mode velocities, $u_1$ and
$u_2$:
$
H=\sum_{i=1,2}\frac{u_i}{2}\int\frac{dx}{2\pi}\left\{
:(\partial_x\eta_i)^2:+:(\partial_x\xi_i)^2:\right\},
$
where the dual fields $\eta_i$, $\xi_i$ are related to the original ones by  linear
transformations
$\vec{\Phi}=\hat{Q}_{\phi}(\eta_1, \eta_2)^T
$,
$
\vec{\Theta}=\hat{Q}_{\theta}(\xi_1, \xi_2)^T
$.
In the case of identical wires, $v_1=v_2=v$, $g_1=g_2=g$, the matrices $\hat{Q}_{\phi,\theta}$
are given by
$Q_{\phi}^{i1}=\frac{1}{\sqrt{2}}
\left[\frac{1/g+\gamma_4+\gamma_2}{g+\gamma_4-\gamma_2}\right]^{1/4}$,
$Q_{\phi}^{22}=-Q_{\phi}^{12}=\frac{1}{\sqrt{2}}
\left[\frac{1/g-\gamma_4-\gamma_2}{g-\gamma_4+\gamma_2}\right]^{1/4}$, and
$Q_{\theta}^{ij}=1/(2Q_{\phi}^{ij})$.
The behavior of the renormalized velocities  $u_1$ and $u_2$ versus the interwire coupling
strength at different values of intrawire interaction constant is shown in the inset in
Fig. \ref{IV-d}.  Each velocity determines the interlevel spacing for a branch of
the plasmon spectrum by the relation $\Delta_i=\pi u_i/L, (i=1,2)$.
 The velocity $u_2$ that renormalizes down by interwire interactions corresponds to the mode,
where the charges of two wires fluctuate in antiphase (the pseudospin mode in the case of
identical wires \cite{Governale,Finkelstein,HS-JvD}). There is a special point in the dependence
of the renormalized velocities on interwire interactions, where $u_1=u_2$. For identical wires,
this point is given by the relation $(\gamma_4+\gamma_2)g=(\gamma_2-\gamma_4)/g$. The farther
away from that point, the lower is the velocity $u_2$.
Moreover, $u_2$ turns to zero when one of the matrices, $M_{\phi}$ or $M_{\theta}$
becomes degenerate.
The latter happens when interwire and intrawire interactions are comparable.
Particularly, in the case of identical wires, and for $\gamma_2=0$, $u_2=0$ is reached for
$\gamma_4=vg$.  The interwire tunneling in
coupled infinite LL's is known to result in an instability either in the $ph$  or
in the $pp$  channel at strong enough interwire interaction
\cite{Finkelstein}.
We expect that one of those two  instabilities  occurs at
the point $u_2=0$, even though the length of the wires is finite. Indeed, the velocity $u_2$
changes the effective length of the wire for the corresponding plasmon mode by the
obvious relation $L\rightarrow L v/u_2$, hence the physics of an infinite LL should be recovered
for $u_2\rightarrow 0$.
Complete analysis of the  instability employing the renormalization group treatment is left for
future publication.

Now we turn to the implications of the above results for the current-voltage characteristics.
We consider here the case, when tunneling between a wire and the next reservoir is much
larger then tunneling between the wires.  Then  the whole
voltage drop takes place between the wires. Increasing the bias voltage,
the energy difference between the two 1P spectral branches $\epsilon_+(k)-\epsilon_-(k)$
increases, which results in the suppression of tunneling matrix  elements $T_{1,r}$ and $T_{2,l}$
($i=1,2$ relates to the LL channel, $\nu=r,l$ relates to the reservoir),
and hence to the suppression of current. However, for a given chirality
$\chi$ ($\chi=1$ for the left, $\chi=-1$ for the right chirality),
that effect can be compensated by  changing the external magnetic field according to
$b=-\chi eV/v_1$ (then both LL channels have the electrochemical potential close to the
left reservoir)  or $b=\chi eV/v_2$ (then both LL channels have the electrochemical potential
close to the right reservoir).  In what follows we consider the case  $b=-eV/v_1$.

The single-particle tunneling current $I$ under the applied voltage $V$ is given by the sum
of the currents through the two parallel LL channels.
In each channel, there are contributions from states with the left and the right
chiralities.
The calculation of the current is outlined in Ref. [\cite{Mahan}], whereas we use
the bosonization formulas for a finite length wire with free boundary conditions to
calculate the fermion correlation function in each channel.\cite{HS-JvD}.
Furthermore, we describe the reservoirs in terms of their density of states with energy
$\epsilon$, $\rho_{r,l}(\epsilon)$.
That is appropriate in the case of two- and three-dimensional reservoirs with large bandwidth,
where the momentum conservation by tunneling does not impose a restriction on energy.
The current-voltage characteristics is obtained as follows,
\begin{eqnarray}
\nonumber
I^{\chi}_{i,r}(eV)=-\frac{16\pi e}{a}|T_{i,r}|^2\sum_{m,n=0}^{\infty}\frac{(b_{i1})_n}{n!}
\frac{(b_{i2})_m}{m!}&&\\
\left\{\theta(-\epsilon_{i,mn}^<)\rho_{r}(\epsilon_{i,mn}^<)
-\theta(\epsilon_{i,mn}^>)\rho_{r}(\epsilon_{i,mn}^>)\right\},&&
\label{Igeneral}
\end{eqnarray}
where
$
b_{i1}=(Q_{\phi}^{i1})^2+(Q_{\theta}^{i1})^2
$,
$
b_{i2}=(Q_{\phi}^{i2})^2+(Q_{\theta}^{i2})^2
$,
$\epsilon_{i,mn}^{\lessgtr}=E(N_i\pm 1,N_j)-E(N_i,N_j)\pm\pi(v_i+n u_1+m u_2)/L-eV$,
and $a$ is a short-range cutoff.
Each term in the sum over $n,m$ describes
the contribution from a plasmon resonance between the two wires to the current.
The tunneling density of states represents a superposition of two sequences of equidistant
$\delta$-peaks with periods $\pi u_1/L$ and $\pi u_2/L$. Both the position and the weight of
each peak is {\it completely defined} by interactions.
The structure of the tunneling density of states has much in common with the structure of
photoluminescence spectra from highly excited nanorings, described in detail in
Ref. [\cite{Shahbazyan}].
Indeed, in both cases the correlations of two finite size LL's determine the relevant
physics with the difference, that it is the four-point correlation function that determines
the luminescence spectra.

A set of IV characteristics for different values of interwire interaction $\gamma_4$ is shown
in Fig. \ref{IV-d}. The shift of the
position of the second step to lower voltages with increasing
$\gamma_4$, which corresponds to the
current through the first excited plasmon state,
demonstrates the facilitation of current by interwire interactions.
For single wall carbon nanotubes, the charging energy is 3 to 6 times the noninteracting
interlevel spacing, \cite{nanotubes,tubes2,nanofin} independently of the nanotube length.
Therefore, the plasmon resonances with $m,n\sim$ 3 --- 6  are expected to be
seen experimentally within a single Coulomb blockade step

For one-dimensional reservoirs, the momentum conservation by tunneling fixes the energy
difference between the modes in the reservoir and in the two-wire system.
The IV curve yields a sequence of peaks.
The positions of the peaks are determined by interference phenomena, which are only partially
controlled by interactions.  Moreover, even applying an external magnetic field,
it is generally impossible to satisfy the momentum conservation and to avoid the diminishing
of tunneling by the bias voltage simultaneously. We conclude that the interaction effects can
hardly be seen if the reservoirs are one-dimensional.

Another effect of interactions is in facilitating plasmon assisted transport at small
voltages and finite temperatures.
For small temperatures,  we obtain $I(T)=I_{1r}(T)+I_{2l}(T)$, where
\begin{equation}
I_{i,\nu}\approx 2 \frac{V|T_{i,\nu}|^2\rho_{\nu}(\epsilon_F)\exp\left(-\frac{v_i\pi}{LT}
\right)}{\left(1-
e^{-\frac{\pi u_1}{LT}}\right)^{b_{i1}}\left(1-e^{-\frac{\pi u_2}{LT}}\right)^{b_{i2}}}.
\label{IT}
\end{equation}
The enhancement of the thermally activated current can be seen by comparison of Eq. (\ref{IT})
with the result for the system without the interwire interactions that reads
$I_{i\nu}\propto \exp\left(\frac{-v_i\pi}{LT}\right)/\left[1-\exp\left(\frac{v_i\pi}{LT}\right)
\right]$.
Due to the smaller interlevel distance
for plasmon excitations in the interacting system (smaller velocity $u_2$), the thermally
activated current is enhanced.

As the singularity $u_2\rightarrow 0$ is approached, the distance between the $m$ and $m+1$
step  in zero-temperature IV characteristics diminishes, the height of each
step grows and formally diverges with $b_{i2}$ at the point $u_2=0$.  The thermally activated
current also grows and diverges as a power-law $(T/u_2)^{b_{i2}}$.
To describe the behavior close to the singularity, the $ph$ and $pp$
hopping operators need to be included in the theoretical treatment, which is beyond the scope
of the present paper.
Beyond the transition, the fluctuations of the fields $(\eta_2,\xi_2)$  are
frozen and do not contribute to the current. The charge fluctuations actually correspond to a
single LL of finite length.  The IV characteristics is
given by  Eq. (\ref{Igeneral}) with $u_2=0, b_{i2}=0$.  The thermally activated current is
suppressed compared to the noninteracting system, since $u_1>v_F$.

The author acknowledges illuminating discussions with M.E. Raikh, who initiated this work.
Discussions with D. Pfannkuche, H. Heyszenau, and S. Kettemann  are gratefully
acknowledged.

\end{document}